\documentclass[preprints,article,accept,pdftex,moreauthors]{Definitions/mdpi} 

\usepackage{longtable}
\usepackage{multirow}

\firstpage{1} 
\pubvolume{1}
\issuenum{1}
\articlenumber{0}
\pubyear{2022}
\copyrightyear{2022}
\datereceived{} 
\dateaccepted{} 
\datepublished{} 
\hreflink{https://doi.org/} 
\Title{Inferring Cultural Landscapes with the Inverse Ising Model}

\TitleCitation{Inferring Cultural Landscapes with the Inverse Ising Model}

\Author{Victor M\o ller Poulsen$^{1}$\orcidA{}, Simon DeDeo$^{1, 2}$*\orcidB{}}

\longauthorlist{yes}

\AuthorNames{Victor M\o ller Poulsen and Simon DeDeo}

\AuthorCitation{Poulsen, V.; DeDeo, S.}

\address{%
$^{1}$ \quad Department of Social and Decision Sciences, Carnegie Mellon University, 5000 Forbes Avenue, Pittsburgh, PA 15213 USA\\
$^{2}$ \quad Santa Fe Institute, 1399 Hyde Park Road, Santa Fe, NM 87501 USA}

\corres{Correspondence: sdedeo@andrew.cmu.edu; Tel.: +1-412-268-3009}

\abstract{The space of possible human cultures is vast, but some cultural configurations are more consistent with cognitive and social constraints than others. This leads to a ``landscape'' of possibilities that our species has explored over millennia of cultural evolution. But what does this fitness landscape, which constrains and guides cultural evolution, look like? The machine-learning algorithms that can answer these questions are typically developed for large-scale datasets. Applications to the sparse, inconsistent, and incomplete data found in the historical record have received less attention, and standard recommendations can lead to bias against marginalized, under-studied, or minority cultures. We show how to adapt the Minimum Probability Flow algorithm and the Inverse Ising model, a physics-inspired workhorse of machine learning, to the challenge. A series of natural extensions---including dynamical estimation of missing data, and cross-validation with regularization---enables reliable reconstruction of the underlying constraints. We demonstrate our methods on a curated subset of the Database of Religious History: records from 407 religious groups throughout human history, ranging from the Bronze Age to the present day. This reveals a complex, rugged, landscape, with both sharp, well-defined peaks where state-endorsed religions tend to concentrate, and diffuse cultural floodplains where evangelical religions, non-state spiritual practices, and mystery religions can be found.}


\begin{document}

\section{Introduction}



If we want to understand the powers and potentials of the human species---the landscape of both what has, and could, be done---we are driven to make comparisons across vast ranges of time and culture. In these cases, data is not only missing, but differentially missing~\cite{clarke1973archaeology}. To analyze, at the same time, a contemporary culture of the digital age, and one that vanished five thousand years ago, requires careful accounting. There is both the intellectual challenge of making best use of what information reaches us, and an ethical imperative to treat long-lost cultures---and marginalized, under-studied, or minority cultures that survive today---on an equal epistemic footing with the dominant, often ``WEIRD''~\cite{henrich2010weirdest} ones, for whom data is both more abundant and more complete.

Doing this well is a challenge. Replacing missing values with ``no'' or ``not present'', for example, is the fallacy of taking absence of evidence for evidence of absence. Replacing them with the median answer, or the best match, from the remainder of the data makes unfamiliar cultures clones of the ones we know. Replacing them with ``a fifty-fifty mixture of present and absent'' is not much better: it attributes the lack of knowledge in the observer to a lack of coherence in the original culture; because we do not know what they did, we assume they did not, either. All these challenges are exacerbated in the ``small data'' limit common in studies of cultural evolution---archives with hundreds of data points, rather than the millions that machine learning algorithms are usually trained and tested on.

This paper addresses the challenge of inferring cultural landscapes in deep time~\cite{smail2007deep,slingerland2017durkheim}. We show how to extend a commonly-used workhorse of machine-learning---the Inverse Ising Problem with Minimum Probability Flow~\cite{mpf}---to the kind of sparse, under-sampled, and potentially biased samples of the historical record. While standard approaches can give misleading answers, we show how a set of carefully constructed modifications and extensions can provide new ways to ask basic questions about the evolution of human culture. We then demonstrate the power, and potential, of cultural landscape construction with an analysis of a curated subset of the Database of Religious History (DRH)~\cite{slingerland2017durkheim,drh2}.


\section{Methods}

The goal of our analysis is the construction of a cultural landscape: a general model of what makes different cultural patterns more or less likely to appear in the course of time. To be more specific, we imagine that we have a set of ``characteristics''---aspects of a culture that we care about, and that can be represented with a binary answer such as {\tt YES} or {\tt NO}, {\tt TRUE} or {\tt FALSE}, {\tt PRESENT} or {\tt ABSENT}, and so on. A particular setting of all the answers is called a configuration, and a landscape model says, for any particular configuration, how likely it is to appear.

Depending on how the experts understand the questions, the landscape derived from it might characterize, on one extreme, the patterns of behavior that could emerge in an individual---or, on the other extreme, the kinds of patterns that entire societies might explore across the span of human history. 

In the case treated here we have cross-cultural data on religious groups in different cultures and time periods from 10,000 BCE to the present day; one characteristic we consider is ``Are supernatural beings believed to mete out punishment?''

A landscape model, could it be found, would be a powerful tool for systematic investigation of how societies navigate practices like these. We might want to know, for example, whether a ``yes'' answer to a belief in punishing Gods makes, say, a ``yes'' answer to ``Does membership in this religious group require sacrifice of children'' more or less likely to be true, all other things being equal.

An answer to such a question would provide important empirical constraints to more fundamental models. One model, for example, might understand child sacrifice as an extreme example of costly signals of devotion in a social context, otherwise disconnected from the metaphysical account the religion provides about God, while another might see the practice as something that could only be conceivable against a particular conceptualization of the relationship between humankind, nature, and the transcendent (see, \emph{e.g.}, Ref.~\cite{child} for discussion). The two models will make different predictions for how the practice co-varies with other characteristics.

Answers to questions like these can not be simply read off from the data, however, because religions with and without a belief in supernatural punishment generically differ on a wide range of characteristics, all of which might impact a violent practice like child sacrifice. The correct answer requires a comparison to a fiducial culture that differs only in one characteristic. The space of configurations expands exponentially, and probing fundamental questions requires knowledge not just of the religions we happen to have observed, but the larger, law-governed landscape of what combinations---including those never observed in human history---are more or less likely.

A landscape model allows us to investigate which features of a religion most strongly couple with others. It would provide insight into how different aspects of a religion bundle together~\cite{norm_bundles}, with a small number of distinct patterns of yes/no answers, as might happen if religions divide into (for example) Axial and pre-Axial types. It would even allow to us identify practices that have yet to emerge---unexplored regions of cultural-evolutionary space. A more prosaic, though no less important, use of a landscape model is to predict missing data. For a long-lost culture, for example, whose metaphysical beliefs are unknown, a landscape model can predict the probabilities of different combinations of epistemic commitments on the basis of its material culture.

\subsection{From Physics to Machine Learning: A Introduction to the Inverse Ising Problem}

Inferring such a fitness landscape from data requires us first to specify the structure of the landscape itself---the spectrum of ways in which it allows one aspect of a pattern to make other aspects more or less likely. In traditional approaches, such as logistic regression, one chooses, ahead of time, a small number of possible effects, based on an explicit model; with a hundred data points, for example, one might try to learn---estimate---three or four regression coefficients. 

When learning a landscape, by contrast, the number of parameters is very large---often comparable to, or even very much larger than, the number of observations~\cite{NEURIPS2021_f197002b}. The particular model we consider in this paper is a very general form of neural network known in the machine learning literature as the ``unrestricted Boltzman machine''~\cite{ackley1985learning}, and (in the physics literature) as the ``inverse Ising problem" (Ackley, 1985).

The inverse Ising model has been applied, with great success, to data ranging from neuroscience~\cite{tkavcik2013simplest,schneidman2006weak}, the immune system~\cite{mora2010maximum}, and the fitness landscapes of HIV~\cite{hiv}, to animal behavior~\cite{bialek2012statistical,daniels2017control}, political polarization and voting behavior~\cite{lee2015statistical,lee2018partisan}, and linguistics~\cite{stephens2010statistical}. It has also been used as a general model of generic complex cultural practices in cultural evolution~\cite{miton}. In one common notation choice, the inverse Ising model says that the probability of observing a configuration $i$ is
\begin{equation}
    p_i = \frac{\exp{E_i(\vec{\theta})}}{Z(\vec{\theta})},\label{boltz}
\end{equation}
where $\vec{\theta}$ are the parameters (to be estimated), $Z(\vec{\theta})$, traditionally called the ``partition function'', is the normalization constant, and the ``energy'', $E_i(\vec{\theta})$, of a particular configuration is given by
\begin{equation}
    E_i(\vec{\theta}) = \sum_{a,b;a>b} J_{ab}\sigma_a\sigma_b + \sum_a h_a\sigma_a,\label{ising}
\end{equation}
where $\sigma_a$ is the truth value of the $a$th entry in configuration $i$; by convention, {\tt YES} is $+1$, and {\tt NO} is $-1$; there are $n(n-1)/2$ of the ``$J$'' parameters (the ``pairwise couplings''), and $n$ of the ``$h$'' parameters (the ``local fields''). 

In general, physicists take the $J$ and $h$ values (or the probability distributions they are drawn from) as given, and try to understand the properties of the resulting distribution~\cite{sherrington1975solvable}. The converse problem, which we consider here, is to infer the ``best fit'' $J$ and $h$ that can predict the observed frequencies of occurance of different configurations in a dataset.

As first noted by E.T. Jaynes~\cite{jaynes1957information}, the form of Eq.~\ref{ising} means that, properly estimated, $p$ is the distribution with maximum entropy that, at the same time, matches the observed means and pairwise correlations; \emph{i.e.}, those found by averaging over all the observed vectors, $\vec{\sigma}_d$, in the dataset $\mathcal{D}$,
\begin{equation}
   \sum_{i} \sigma_a p(i) = \frac{1}{|\mathcal{D}|}\sum_{d \in \mathcal{D}} \sigma_{a,d} \hspace{0.5cm} \textrm{and} \hspace{0.5cm}   \sum_{i} \sigma_a\sigma_b p(i) = \frac{1}{|\mathcal{D}|}\sum_{d \in \mathcal{D}} \sigma_{a,d}\sigma_{b,d}  \label{constraints2}
\end{equation}
Such models embody a kind of inverted form of Occam's Razor: make the model just sophisticated enough to explain only the least complicated features of the data to hand, leaving everything else maximally undetermined. Surprisingly enough, this  works: as has been repeatedly discovered, higher-order correlations often ``come along for the ride'', emerging spontaneously when the pairwise constraints, Eq.~\ref{constraints2} are satisfied~\cite{schneidman2006weak,stephens2011searching,chris}. Despite its simplicity, Eq.~\ref{ising} can capture a great deal of the real variability in complex systems, and many of the most celebrated successes of machine learning are, at heart, adaptations of this insight~\cite{inverse}.

\subsection{Minimum Probability Flow}

Finding the values of $J$ and $h$ that satisfy Eq.~\ref{constraints2} is exponentially hard, because it requires averaging over all $2^n$ configurations in the probability distribution Eq.~\ref{boltz}. We can rephrase the problem, however, as trying to find the Ising-model distribution, $p_i(J_{ab}, h_a)$, that best fits the true (or ``data'') distribution, $p_i$. where the ``best fit'' is the one that minimizes the Kullback-Leibler divergence,
\begin{equation}
    K(\vec{\theta}) = \sum_{i \in \mathcal{C}} p_i \log_2 \frac{p_i}{p_i(\vec{\theta})}, \label{cross}
\end{equation}
where $\mathcal{C}$ is the (exponentially large) set of all $2^n$ configurations. The $\vec{\theta}$ that minimizes Eq.~\ref{cross} produces a $p_i(\vec{\theta})$ that is minimally-distinguishable, in a basic information-theoretic sense, from the true distribution $p_i$.


Minimizing Eq.~\ref{cross} directly, however, still requires multiple sums over $\mathcal{C}$. The insight of MPF~\cite{mpf} is that, given a collection of observed configurations, $\mathcal{D}$, Eq.~\ref{cross} can be approximated by minimizing the ``probability flow''. When a parameter choice $\vec{\theta}$ is a poor match to the data, probability tends to flow ``away'' from data states to non-data states. Up to constant factors, we can approximate Eq.~\ref{cross} as
\begin{equation}
    K(\theta) = \sum_{j\in D} \left( \sum_{i\in\mathcal{N} \notin \mathcal{D}} \Gamma_{ij}(\vec{\theta}) \right),\label{final}
\end{equation}
where $\Gamma_{ij}(\vec{\theta})$ is the rate of flow from state $j$ to state $i$ for parameter choice $(\vec{\theta})$, and $\mathcal{N}$ is a set of ``neighbouring'' non-data configurations. Minimizing Eq.~\ref{final} is a tractable task; in contrast to Eq.~\ref{cross}, the sums are no longer over $\mathcal{C}$, but a radically smaller set of observed data, $\mathcal{D}$, and a well-chosen $\mathcal{N}$. MPF is related to a basic method in machine learning, Contrastive Divergence~\cite{hinton2002training}, with the principle advantage, for our purposes, of having a well-defined, epistemically principled, objective function.


\subsection{Improvements and Extensions to the MPF Algorithm}

In this section, we present a series of improvements and extensions to the basic MPF algorithm. These include both apparently minor, but critical, variations on the basic algorithm, and a new extension and derivation. We are particularly grateful to the authors of {\tt ConIII}~\cite{coniii}, whose implementation, and clear discussion, of MPF enabled us to debug and test our own code.

Sections~\ref{nn} and~\ref{reg_constraint} present a pair of improvements to the basic algorithm; these provide significant boosts in performance and accuracy on sparse social and cultural data. Section~\ref{inconsist} shows how to handle inconsistencies between different observers (or inconsistencies within the same observer), and Section~\ref{time} shows how the same tools also allow us to account for uneven sampling in time or space. Finally, Section~\ref{missing_data} describes a novel extension to the MPF algorithm, Partial-MPF, which enables us to handle missing data in a principled fashion.

\subsubsection{Nearest-Neighbour Sampling}
\label{nn}

In the original version of the MPF algorithm, flow is computed from the observed configurations (``data states'') to a subset of other configurations, explicitly excluding flow into any other data states. It is equally valid, under the MPF approximation, to allow flow into states that do appear elsewhere in the data; this can be seen at line A-6 of Ref.~\cite{mpf}, where you can interchange the order of the derivative and the summation. This alternative choice is the default under {\tt ConIII}. 

Our experiments find that the alternative choice provides greatly improved out-of-sample performance, because the exclusion biases the algorithm against configurations near a metastable peak. With this change in hand, the function to be minimized is
\begin{equation}
    K(\theta) = \sum_{j\in D} \left( \sum_{i \in \mathcal{N}(j)} \Gamma_{ij}(\vec{\theta}) \right),\label{final2}
\end{equation}
A natural choice is to set $\mathcal{N}(j)$ to include states within a certain Hamming distance of $j$; the original MPF paper considered states that differed from the data state at one position, \emph{i.e.}, $\mathcal{N}_1(j)$; we also consider a strategy that uses states up to two ($\mathcal{N}_2(j)$) Hamming units away. Since $|\mathcal{N}(j)|$ is the same for all $j$, this provides equal weighting to all data states. (It is also possible to consider randomly chosen neighbours; however, this tends to give significantly decreased performance; the MPF algorithm performs best when it is allowed to focus on reasonably nearby variations from the observations.)

\subsubsection{Regularization Constraint}
\label{reg_constraint}

Minimizing Eq.~\ref{final2} is equivalent to (attempting to) maximize the posterior log-probability of the data given the model. A proper Bayesian analysis, however, should include not just the posterior, but a prior over the parameters themselves,
\begin{equation}
    K^\prime(\theta) = K(\theta) - \lambda|\mathcal{D}||\mathcal{N}|\log{P(\vec{\theta})},\label{final_sparsity}
\end{equation}
where $\lambda$ is a constant, and $P(\vec{\theta})$ is the probability of a particular choice for $J$ and $h$.

It is natural to choose $P(\vec{\theta})$ so that, all other things being equal, smaller values are preferred; this is sometimes known as a regularization penalty, which often provides significant benefits to out-of-sample prediction~\cite{bickel2006regularization}. Without regularization, models tend to overfit, producing unreasonably low probabilities for configurations that happen not to appear in the data.

If we assume that $J$ and $h$ are distributed as a Gaussian---what is sometimes known as the $L2$-norm---then we have
\begin{equation}
    K^\prime(\theta) = K(\theta) - \lambda|\mathcal{D}||\mathcal{N}|\sum_{k=1}^{N_p} \frac{\theta_k^2}{2},\label{final_sparsity_2}
\end{equation}
where the value of $\lambda$ encodes the variance of the Gaussian; larger $\lambda$ corresponds to smaller variance.

The optimal choice for $\lambda$ depends on $P(\vec{\theta})$---which is, in general, unknown. It can be estimated, however, by cross-validation: if there are $m$ datapoints, fit the data using $m-k$ datapoints (the training set), and compute the log-likelihood for the remaining $k$ datapoints (the test set). In this paper, we take $k$ equal to one, \emph{i.e.}, leave-one-out cross-validation. Repeating this for all possible choices of the left-out observation, and then averaging the result, allows us to estimate the performance of the fit as a function of $\lambda$.

\subsubsection{Inconsistent data}
\label{inconsist}

In some case---for example, in about $17\%$ of religious groups in the DRH data used below---we have inconsistent coding, where multiple, incompatible answers exist for the same configuration. This can emerge when different observers interpret a question, or evidence, in different ways, or have different examples in mind. In the DRH, it most commonly appears when the same observer flags a feature as less straight-forward than it appears; for example, ``Iban traditional religion" is inconsistently coded for whether the religion had scriptures, with the coder citing it as a ``borderline case" and answering both ``yes", and ``no". Another example is ``Unitarian Universalism'' (UU), where the same observer coded belief in afterlife as both ``yes", and ``no", noting that some UUs do, and some do not, believe in an afterlife. A proper accounting of the landscape ought to allow for both.

To make explicit use of inconsistent data requires an error model, and there are two natural choices. Consider, as an example, two observers who provide inconsistent answers, for the same system, to three binary questions: $j_{1}$ gives $\{1, 1, 0\}$, while $j_{2}$ gives $\{1, 0, 1\}$. If we assume that, for each observer, their best answer to one question is dependent upon all the others, we can include both records, with a weighting term, $w_j$, that captures the epistemic uncertainty 
\begin{equation}
    K(\theta) = \sum_{j\in D} \left( \sum_{i \in \mathcal{N}(j)} w_j \Gamma_{ij} \right), \label{weighting}
\end{equation}
where $w_{j_1}=w_{j_2}=1/2$. Alternatively, one can take inconsistencies as evidence of uncertainty question-by-question--the ``independent'' model. Then we interpret the observations $j_1$ and $j_2$ as indicating that observers are, in general, uncertain about the answers to questions two and three, with independent probabilities of ``yes'' for each $1/2$. In this case, one includes not only the observed records ($r_{1} = \{1, 1, 0\}$, $r_{2} = \{1, 0, 1\}$) but also the unreported combinations $r_{3} = \{1, 1, 1\}$ and $r_{4} = \{1, 0, 0\}$, each with weight $1/4$. 

Both choices imply that differences between observers trace back, not to uncertainty about a fixed reality, but rather to fluidity in the practices themselves, where both answers are equally valid depending on the details of time and place. The examples presented here are the most common form of inconsistency, and this argues in favor of the independent model. 

\subsubsection{Correcting for non-uniform weighting across time and space}
\label{time}

Cultural data is often unevenly sampled. We have more examples from the present than the distant past; more from the developed world than from the developing world; more from dominant cultures in a region than marginalized or minority ones.

This can lead to bias in our landscape estimation. If we have, for example, twenty observations from cultures of Type A (the ``contemporary developed world'' sample), and only ten observations from cultures of Type B (the ``understudied'', or ``minority'', sample), then a naive use of the data would tend to lead to landscapes that made Type A cultures look more stable than Type B cultures, and would produce accounts of the interlocking constraints that made Type A cultures look more natural than Type B cultures. 

Often, however, we will know from archival records or field reports that groups exist, even if we know nothing about them, which allows us to estimate the sampling bias. With such an estimate in hand, Eq.~\ref{weighting} allows us to re-weight observations to compensate.


\subsubsection{Partial-MPF: Accounting for missing data}
\label{missing_data}

Handling missing data is a challenge. Consider an observation such as the following,
\begin{eqnarray}
    j = \{1, 0, X, X\},
\end{eqnarray}
where answers to the last two questions are not provided. The function that MPF minimizes, Eq.~\ref{final}, can only be calculated for fully-specified data, and so a natural response is to do data imputation: for example, replacing missing answers with the most common responses for that question in the remainder of the data. 

While naive imputation methods are often suggested in machine learning tutorials, they are, in the final analysis, an epistemic fallacy: they replace what is unknown by what is known, and assume that what hasn't be seen looks like what has. In qualitative work, such a fallacy would be obvious. An archaeologist would not suggest, for example, that the metaphysical beliefs of a long vanished civilization should match the median beliefs of civilizations today.

A better way to solve this problem, which we refer to as ``Partial-MPF'', is to dynamically infer the missing data from the best estimates of the parameters $\vec{\theta}$; \emph{i.e.}, to work not with a particular completion for $j$, but a distribution over, in this case, the four possible values, $j_1$, $\{1, 0, 0, 0\}$, $j_2$, $\{1, 0, 0, 1\}$, $j_3$, $\{1, 0, 1, 0\}$, and $j_4$, $\{1, 0, 1, 1\}$, found using Eq.~\ref{boltz}.

When the amount of missing data is small (in practice, less than ten missing values per configuration), the distribution can be computed exactly. For an observation with $m$ missing values, we expand the observation into the $2^m$ different combinations, compute the weights, $w_j(\vec{\theta})$, for each combination, and combine them together as in Eq.~\ref{weighting}. This is somewhat like the ``expectation-maximization'' step suggested by Ref.~\cite{battaglino2014minimum} for missing data, but with probabilistic weightings that preserve continuity in the derivative.

Doing this correctly requires care, and there are three alterations we have to make to the basic algorithm. First, we must update the weights $w_j(\vec{\theta})$ as we move through parameter space. Second, because the weights depend on $\vec{\theta}$, this changes the form of the derivative $dK(\theta)/d\vec{\theta}$. Third, while the inference of the missing data is exact, $K(\vec{\theta})$ is still only an approximation, and so minimizing $K(\vec{\theta})$ will be in slight tension with the inference step. As we will see in Section~\ref{missing}, this is not a show-stopper, and our treatment of missing data is, in practice, much more effective than standard alternatives.

\section{Data}

Our case study draws on data from the Database of Religious History\footnote{\url{http://religiondatabase.org}} (DRH)~\cite{slingerland2017durkheim,drh2}. The DRH, an ongoing project based at the University of British Columbia, includes a peer-reviewed collection of information about religious groups in both the contemporary, historical, and archaeological record, in the form of coded answers to standardized question sets (``polls'', in the DRH)~\cite{slingerland2017durkheim, slingerland2020coding,SPICER2022100073}. 

The DRH is organized hierarchically, such that some ``super''-questions (e.g. ``Is a spirit-body distinction present?") have sub-questions (e.g. ``Is spirit-mind conceived of as having qualitatively different powers or properties than other body parts?"), and even sub-sub-questions. For this case study, we limit ourselves to super-questions, since sub-questions are contingent on answers to super-questions. This limits the number of (related) questions from $1133$ to $171$. The majority of the questions are binary questions, and so are a natural fit to the Inverse Ising method. When we limit ourselves to questions that ask for binary answers, this further limits the number of questions from $171$ to $149$, and the number of records from $838$ to $835$.

The DRH is under continuous development. In this preliminary analysis, intended to demonstrate the methods and the basic ideas behind landscape construction, we focus on a subset of $20$ questions, and do not correct for potentially uneven sampling of groups by time or place. We start by selecting the questions with the fewest unanswered questions across records, and then select all records (\emph{i.e.}, religious groups) that have five or fewer missing answers. Additionally, selecting only civilizations from the ``group" poll~\cite{slingerland2020coding}, leaves us with a final data set of $407$ civilizations, and we give this set to the Partial-MPF algorithm for the final stage of actually inferring the parameters. See Appendix~\ref{question_table} for the full list of questions, and Appendix~\ref{observed_religions} and~\ref{unobserved_religions} for all religious groups in our curated dataset

\section{Results: Simulations}
\label{valid}

We first present the results of simulations; these confirm that our extensions to the basic MPF algorithm provide critically important improvements to the quality of the fit. To do this, we create large numbers of ``imaginary'' landscapes, where the underlying parameters have statistics similar to those observed in the real world. We take $n$, the number of {\tt YES}/{\tt NO} questions, equal to twenty, and we draw the parameters $J_{ab}$ and $h_a$ from a Gaussian distribution. We then simulate data as draws from this underlying distribution, using the Metropolis-Hastings algorithm, altering it in different ways to take into account how real-world data is distorted by the data-gathering process.

With these simulated data sets in hand, we use our different extensions to the MPF algorithm to attempt to infer the underlying true parameters. We quantify the performance of our algorithms by direct calculation of the Kullback-Leibler divergence of the inferred distribution (corresponding to inferred parameters $\hat{J}_{ab}$ and $\hat{h}_a$) from the true distribution (which, in our simulations, is known---just the distribution produced by the original $J_{ab}$ and $h_a$),
\begin{equation}
    \mathrm{KL} = \sum_{i \in \mathcal{C}} p_i(J_{ab}, h_a) \log \frac{p_i(J_{ab}, h_a)}{p_i(\hat{J}_{ab}, \hat{h}_a)}.
\end{equation}
When KL is close to zero, the inferred distribution is hard to distinguish from the true distribution. KL has a number of useful properties that allows it to play the role of ``mean squared error'' for probability distributions~\cite{kline2005revisiting}, quantifying the relative error in reconstruction and prioritizing accurate reconstruction of the more common states.

In general, reconstruction performance will depend upon the parameters of the distribution from which the test values $J_{ab}$ and $h_a$ are drawn. For our particular case of $N=20$, we choose this to be a Gaussian with mean zero, and $\sigma$ that ranges between $0.01$ and $1.0$. 

When $\sigma$ is small, the constraints are very weak and we are in a near-random or ``dispersed'' regime. As $\sigma$ gets larger, we enter what we call the ``ordered'' regime up to $\sigma$ of approximately $0.25$, where constraints are strong enough to produce peaks where data tends to cluster; practically speaking, this is where most real-world systems, including the DRH, tend to be found. For completeness, we consider yet larger $\sigma$ values: going above $0.25$ we enter the ``near critical'' regime, where these peaks become sufficiently strong to produce large-scale order, and finally, what we call the ``critical'' regime, above $0.5$, where the distribution is near, or past, the spin glass phase transition.

\subsection{Regularization and Cross-validation greatly improve performance}
\label{regularization}

\begin{figure}[ht]
    \centering
    \includegraphics[width=0.7\textwidth]{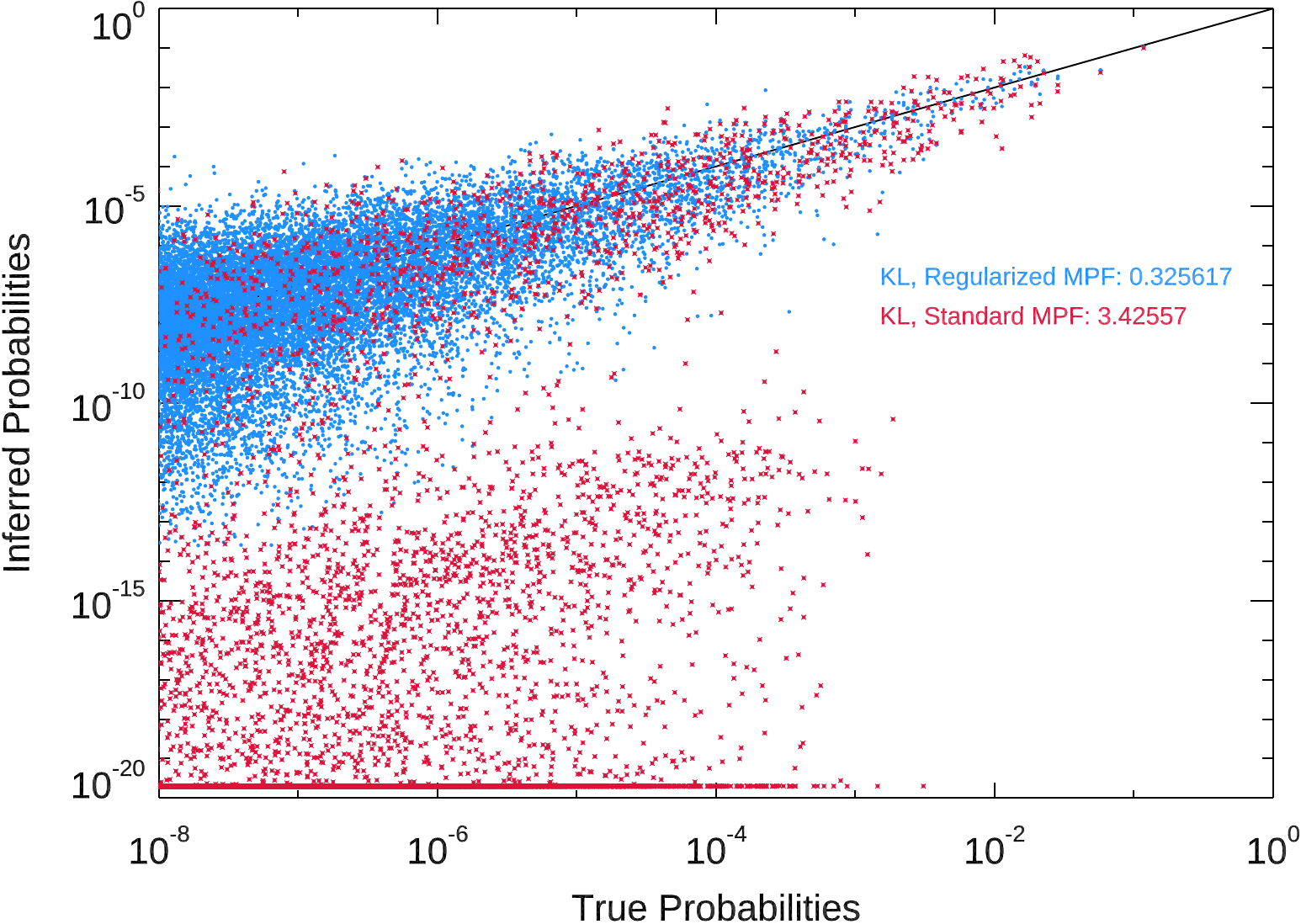}
    \caption{Regularization corrects for overfitting. A sample reconstruction of the $2^{20}$ ($\approx$1 million) probabilities for a landscape, based on $256$ datapoints. Without the regularization constraint (red ``$\times$'' points), the model underestimates the probabilities of some reasonably common configurations. The effect is largely controlled when using regularization with cross-validation (blue points).}
    \label{regularization_plot}
\end{figure}
Regularization using the $\lambda$ parameter significantly improves our ability to estimate the underlying landscape, making reliable extraction possible with very small amounts of data. An example is shown in Figure~\ref{regularization_plot}, where we take a particular simulated dataset (with $\sigma$ equal to $0.5$), and compare the probabilities estimated using the baseline MPF (\emph{i.e.}, without regularization), to our regularization method where $\lambda$ is estimated using leave-one-out cross-validation.

The regularized model is not only better at estimating the probability of the peaks of the landscape (the more likely, high-probability configurations), it also avoids overfitting to less common configurations. Standard MPF, by contrast, can sometimes recover very large values for the $J_{ab}$ parameters, leading it to underestimate the vast majority of the less-likely configurations ($p$ less than $10^{-2}$). For Standard MPF, sometimes, what has not been seen is not just less likely, but effectively impossible.

\begin{table}[ht]
    \centering
    \begin{tabular}{l|c|c|c}
$\beta$ range & Optimal KL & KL with CV & Standard MPF \\ \hline
& \multicolumn{3}{c}{$n$=20, 128 points} \\ \hline
0.01--0.125 (dispersed) & 0.22 & 0.23 & 1.2 \\ 
0.125--0.25 (ordered) & 0.55 & 0.56 & 2.3 \\ 
0.25--0.5 (near-critical) & 0.62 & 0.63 & 19.4 \\ 
0.5--1.0 (critical) & 0.50 & 0.54 & 9.5 \\ 
    \end{tabular}
    \caption{Cross-validation can recover near-optimal sparsity parameters. Without sparsity, MPF consistently overfits to observed data. Reconstruction with 20 nodes (210 parameters), and 128 data points (\emph{i.e.}, the undersampled regime). The more computationally-expensive $\mathcal{N}_2$ strategy does not improve significantly over the simpler $\mathcal{N}_1$.}
    \label{sparsity_nn1}
\end{table}
\begin{table}
    \centering
\begin{tabular}{l|c|c|c|c|c}
$\beta$ range & Ideal & \multicolumn{4}{c}{Biased Sample} \\ \hline
 &  & \multicolumn{2}{c|}{KL} & \multicolumn{2}{c}{Bias against minority} \\ \hline
 & & Corrected & Naive & Corrected & Naive \\ \hline
0.01--0.125 (dispersed) &  0.13 & 0.16 & 0.16 & -0.2\% & -14\% \\ 
0.125--0.25 (ordered) &  0.34 & 0.45 & 0.46 & -0.1\% & -40\% \\ 
0.25--0.5 (near-critical) &  0.43 & 0.55 & 0.60 & 9.6\% & -51\% \\ 
0.5--1.0 (critical) &  0.48 & 0.57 & 0.71 & 0.1\% & -65\% \\ 
\end{tabular}
    \caption{Reweighting observations can correct for sample bias.}
    \label{bias_correction}
\end{table}

Table~\ref{sparsity_nn1} shows that regularization makes reconstruction possible even in the radically under-sampled regime where the number of parameters (here, 210, for $n=20$) exceeds the amount of data (here, 128 observations), and cross-validation leads to near-optimal results.

\subsection{Re-weighting can correct for sampling bias}
\label{bias}

To study bias correction, we simulate multiples examples of a biased sampling process. First, we construct landscapes (for a variety of $\beta$ values) where answers to one of the questions is split, evenly, between {\tt YES} (the ``Type A'' groups) and {\tt NO} (the ``Type B'' groups). We then create two samples: a full sample of 256 observations, and a biased data sample, with 128 observations of Type A groups, but only 64 observations of Type B groups. This simulates an extreme form of bias, where the dominant Type A cultures are over-sampled by a factor of 2:1

We then compare the reconstruction performance in three conditions: the ideal case, with 256 observations, the naive-biased case, where parameters are learned from the biased sample, and the re-weighting case, where we implement the weighting prescription of Section~\ref{time}. We measure both the KL divergence, and the average log-odds bias against the Type B groups, defined as
\begin{equation}
\mathrm{Bias}=\exp\left(\left\langle \log{\frac{p_B}{p_A}} \right\rangle\right) - 1,
\end{equation}
where $p_B$ is the model's predicted probability of Type B groups, $p_A$ (equal to $1-p_B$) is the predicted probability Type A groups, and the average is taken over multiple simulations in a $\beta$ range. The true value, by construction, is $p_A$ equal to $p_B$, and negative values indicate bias against the minority cultures.

Table~\ref{bias_correction} shows the results; even at 2:1 levels of bias, our methods can achieve high reconstruction accuracy without inappropriately biasing the underlying landscape in favor of the dominant cultures.

\subsection{Partially-observed data can be consistently integrated into inference}
\label{missing}

\begin{figure}[ht]
    \centering
    \includegraphics[width=0.75\textwidth]{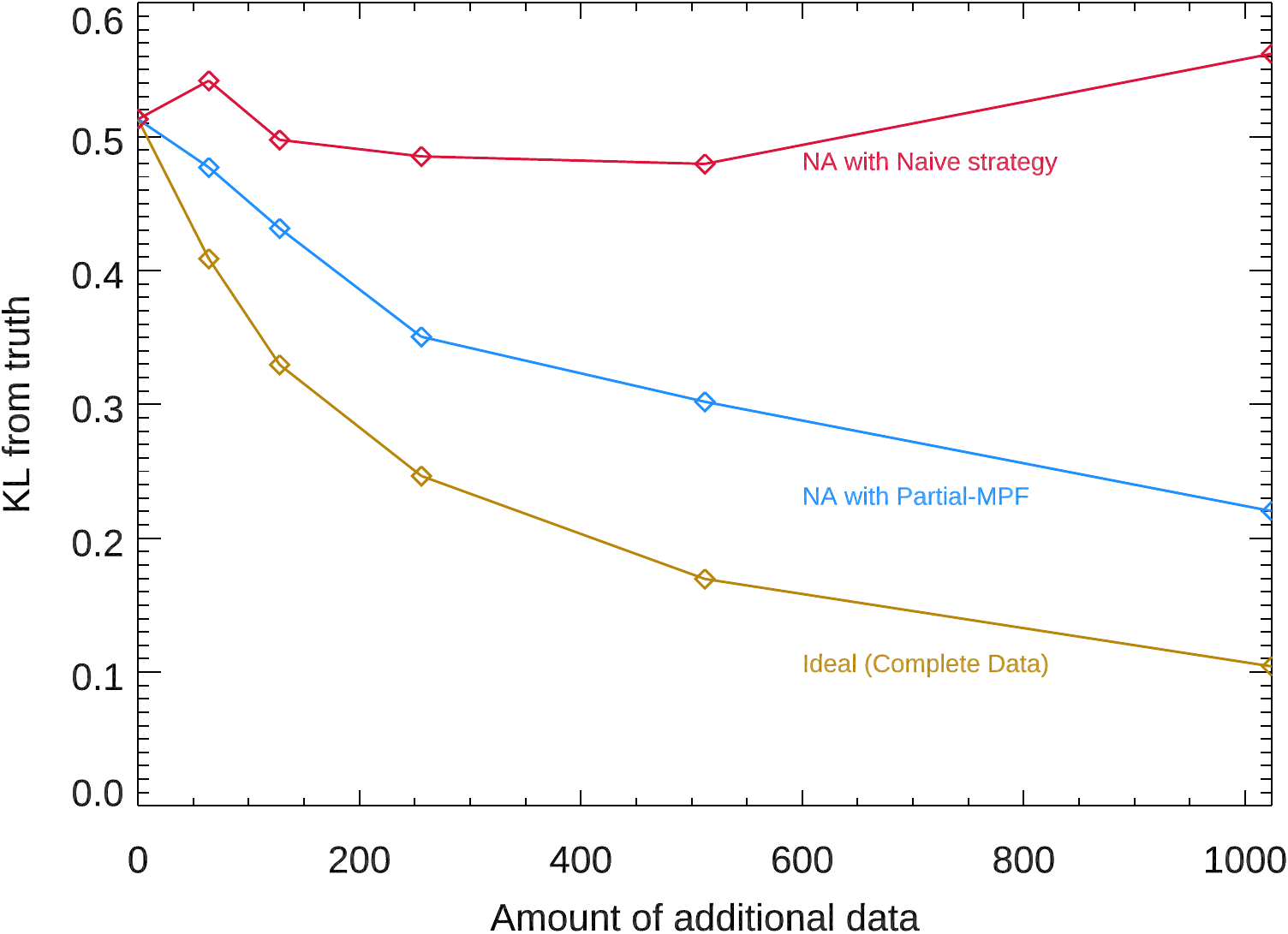}
    \caption{A example of how Partial-MPF adapts the baseline MPF algorithm to make use of partial data. As more, but incomplete, data is added, the Partial-MPF fit (blue line) continues to improve, though not as fast as when data is complete (yellow line). By contrast, the naive strategy often makes the fit worse, because imputation destroys implicit correlations.}
    \label{na_figure}
\end{figure}

To test the performance of Partial-MPF, we consider a scenario where we have a certain amount of complete data, and then add in new partially-observed data. Figure~\ref{na_figure} shows an example of how this works in practice for a single simulated system.  We begin with 128 data points, and then add increasing amounts of data that is 25\% incomplete (a random selection of five of the twenty features are blanked out.) We compare our method to a common ``naive'' choice of taking missing variables to have the most commonly observed value in the remainder. 

The three lines show how fit quality changes as (1) more fully-observed data is added (the ideal case); (2) partially-observed data is added, and integrated in using the Partial-MPF strategy; and (3) partially-observed data is added, using the naive strategy. While Partial-MPF is able to make good use of the data to improve the fit (the KL from truth declines), additional (noisy) data appears to harm the quality of the naive fit beyond a certain point. Table~\ref{partial_wins} shows the average results in different regimes; the same pattern is observed.

\begin{table}
    \centering
\begin{tabular}{l|c|c|c|c}
$\beta$ range & 128 full & \multicolumn{2}{c|}{128 full + 128 partial} & 256 full \\ \hline
& & Partial-MPF & Naive & \\ \hline
0.01--0.125 (dispersed) &  0.22 & 0.17 & 0.24 & 0.15 \\ 
0.125--0.25 (ordered) &  0.56 & 0.44 & 0.58 & 0.34 \\ 
0.25--0.5 (near-critical) &  0.63 & 0.50 & 0.82 & 0.40 \\ 
0.5--1.0 (critical) &  0.54 & 0.43 & 0.87 & 0.38 \\ 
\end{tabular}
    \caption{Using Partial-MPF to reconstruct landscapes in the presence of partially-observed data. While the ``naive'' strategy actually decreases the quality of the fit, Partial-MPF enables efficient use of partial observations to improve knowledge of the landscape.}
    \label{partial_wins}
\end{table}

\section{Results: The Database of Religious History}
\label{drh-data}

We present our empirical results in four parts. First, in Section~\ref{params_section}, we look at the values of the inferred parameters. Doing this shows how to read the underlying ``logic'' of the landscape: the key interactions that combine together to make some configurations more consistent with constraints than others.

We then look at the landscape itself. In Section~\ref{predict}, we show how it can be used to inform hypotheses in cases where data is inconsistent or missing; we take, as an example, the case of a cult in the ancient Mediterranean.

In Section~\ref{visualize}, we show how to visualize the large-scale structure of the landscape---the topography of ``peaks'' (concentrated regions where religions tend to cluster), ``valleys'' (where underlying constraints make traditions harder to sustain), and ``floodplains'' (areas of configuration space where constraints are weaker, favoring diversity and variation). Finally, in Section~\ref{focal}, we show how to analyze the local neighbourhood of a configuration, which gives us a new window onto the question of cultural evolution over time.

\subsection{Parameter Interpretation and Landscape Logic}
\label{params_section}

\begin{figure}[ht]
    \centering
    \includegraphics[width=\textwidth]{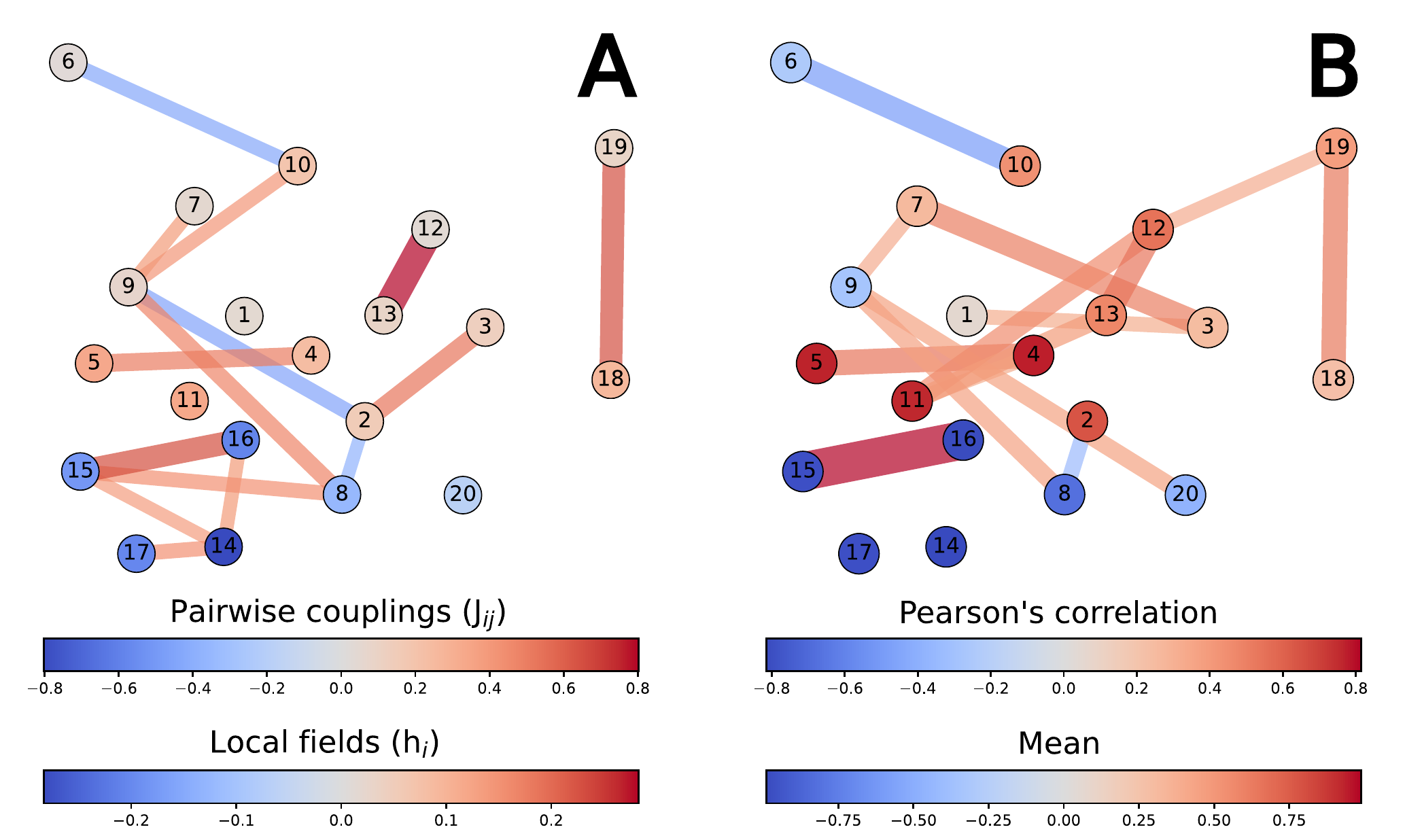}
\caption{The logic of the cultural landscape (left), compared to the surface-level correlations (right). Nodes represent questions; see Table~\ref{question_table} for the question text. Left: edges represent the fifteen strongest pairwise couplings ($J_{ij}$) between questions, as inferred by Partial-MPF; nodes (questions) are colored by the value of the local fields $h_i$. Right: edges represent the fifteen strongest Pearson correlations; nodes are colored by the observed mean. Node placement (layout) is explained in the Appendix~\ref{net_layout}.} 
    \label{param_plot}
\end{figure}

Figure~\ref{param_plot} provides a simple overview of the logic of the cultural landscape derived from the DRH. This compares the underlying parameters of the Inverse Ising model (the $J_{ij}$ and $h_i$ terms), inferred by Partial-MPF, to the surface-level, observed correlations in the data. 

In some cases, the surface-level correlations are a good guide to the underlying logic. Our model suggests that, for example, the observed correlation between between small-scale (18) and large-scale (19) rituals is most naturally explained, at this resolution, by an underlying sympathetic (\emph{i.e.}, $J_{18,19}$ positive) pairwise constraint. Similarly, the ``big Gods''~\cite{norenzayan2013big} pairing of supernatural monitoring (12) and supernatural punishment (13) is both a strong surface-level feature, and a core part of the landscape logic.

Much of the surface-level structure that we observe, however, turns out to be an emergent property of more complex relationships in the underlying parameters. The model suggests, for example, that a strong surface-level correlation between monumental architecture (3) and special treatment for corpses (7) can be explained away by mediation through other variables. Grave goods (9) is another example: it is rare in the observed data, but the local field for this feature is slightly positive, suggesting that there is nothing inherently difficult about maintaining a grave-good tradition. Instead, the practice is disfavored because of how it interacts with, for example, the keeping of written scriptures (2). Our model also reveals an underlying logic that links interactions among an ``extreme'' set of practices (castration (14), adult sacrifice (15), child sacrifice (16), grave-co-sacrifices (8), and suicide (17)).

\subsection{Hypothesising the Unknown}
\label{predict}

Landscape models enable us to predict unknown data: given partial information about a group, Eq.~\ref{boltz} allows us to conjecture about how the constraints, inferred from other systems, would interact in the particular case at hand. Cases with genuine expert disagreement, and cases where features of religious cultures are unknown due to the ravages of time, are the most exciting to analyze in this way.


As a particularly compelling example, consider the ``Archaic Spartan Cults" (800 BCE---500 BCE). For these precursors to the Spartan state, both the presence of child sacrifice and small-scale rituals have been coded by the DRH expert as ``unknown to the field''. The inferred parameters, along with what \emph{is} known about the Spartans, however, provide the following degrees of belief in the four combinations,
\begin{equation}
        \begin{tabular}{c|c|c}
            & Small-scale Ritual & No Small-scale Ritual \\ \hline
           Child Sacrifice  & 1.3\% & 0.3\% \\
           No Child Sacrifice  & {\bf 78.6\%} & 19.9\% \\
        \end{tabular} \nonumber
\end{equation}
The model is nearly 99\% certain that the Cults did not practice child sacrifice. In this case, the known absence of both castration and adult sacrifice, both of which have sympathetic links with child sacrifice in the underlying model, are sources of evidence against the proposition (see Figure~\ref{param_plot}A). 

The model is also reasonably confident about the presence of small-scale rituals; here, emergent constraints such as the strong pairwise coupling to the presence of large-scale rituals, which the Spartan Cults are known to have had, tilt the balance in favor of small-scale ritual. The judgement is less certain, however. The power of the Inverse Ising model is seen here not just in its recognition of common patterns, but in how it parses out of the evidentary value of different pieces of information.

\subsection{The Landscape of Religious Culture}
\label{visualize}

Our parameters imply a probability distribution over $2^{20}$ possible configurations: a landscape with peaks (small groups of high-probability configurations), and valleys (areas of low-probability configurations). As we shall see, landscapes also can include wider ``floodplains''---more widely dispersed collections of configurations that are relatively, and relatively equally, probable.

It is difficult, however, to view all the configurations at the same time: placing \emph{all} the points of a twenty-dimensional hypercube on a two-dimensional plot makes it very hard to see which configurations are close (and, \emph{e.g.}, part of a connected plateau) vs.\ far (\emph{e.g.}, two well-separated peaks). 

\begin{figure}[ht]
    \centering
    \includegraphics[width=\textwidth]{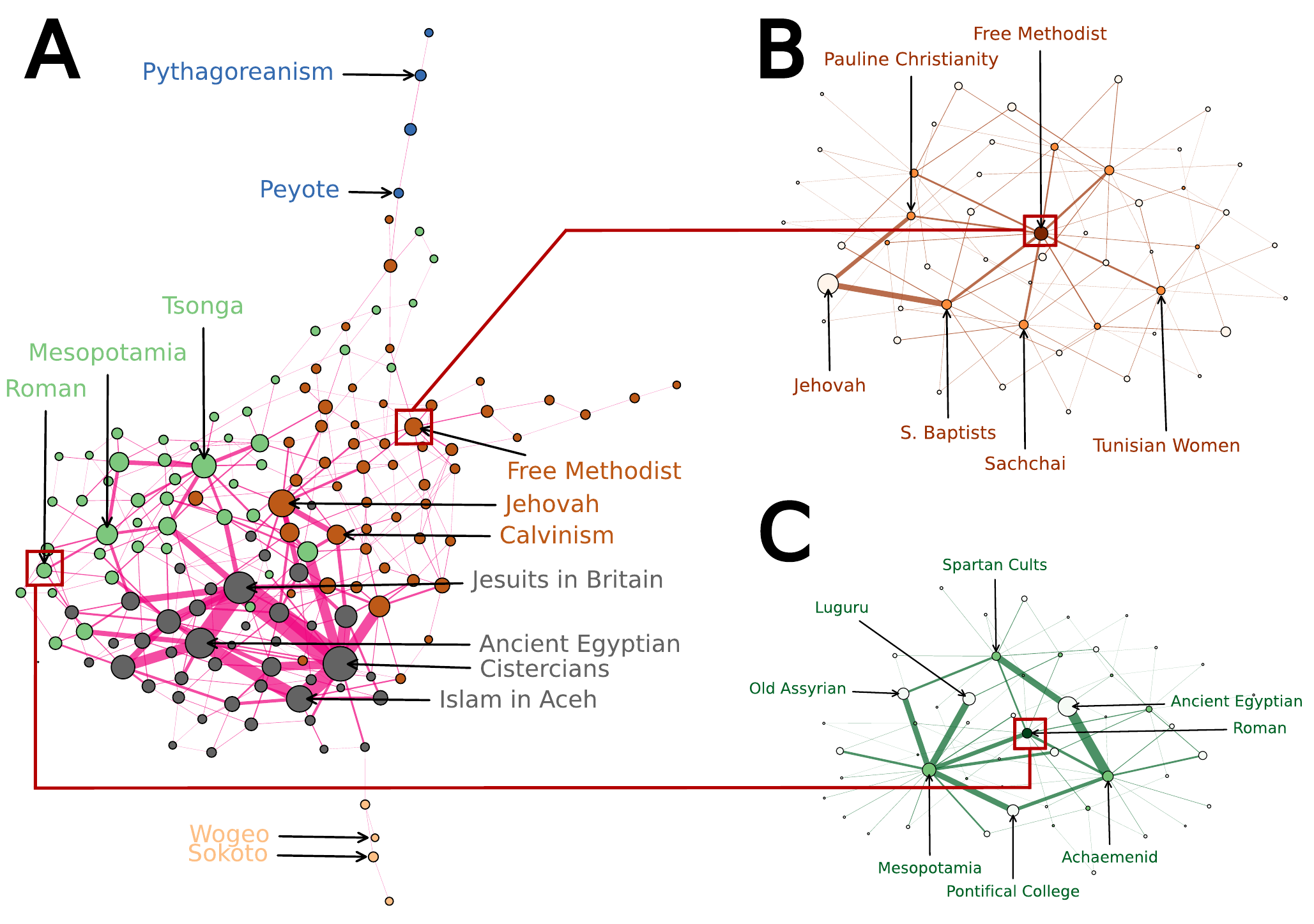}
    \caption{Figure A (left) shows the 150 configurations that have the highest probability mass according to our model. We only show edges between configurations (nodes) that are immediate neighbors (separated by 1 Hamming distance). Nodes are scaled by the probability mass assigned to each configuration, and edges are scaled by the product of the probability mass of the nodes that they connect. Colors assigned to each of five groups obtained from community detection (see Appendix Section~\ref{comm_detect}). Figure B shows the 50 most probable configurations in the local neighborhood of the Free Methodist Church, while figure C shows the 50 most probable configurations in the local neighborhood of the Roman Imperial Cult. Scaling of nodes and edges the same as in figure A. Node-color in both cases (B, C) based on Hamming distance from the central node (e.g. Free Methodist Church dark orange, 1-Hamming distance light orange, and 2-Hamming distance white). In all cases (A, B, C), the layout is determined by a force-directed placement algorithm~\cite{fruchterman1991graph} as implemented in Graphviz~\cite{ellson2001graphviz}. For more on the layout approach see section~\ref{net_layout}.}
    \label{config_networks}
\end{figure}

One way to approach this problem is to start with the topography of the peaks. In Figure~\ref{config_networks}, we represent the 150 most probable configurations as a network, where configurations that differ in one answer are connected by an edge, and the nodes are arranged to best represent distances; roughly speaking, configurations that differ in more answers are further apart (see Section~\ref{net_layout} for details). The configurations shown in the network contain $46\%$ of the total probability mass, and can provide an overview of the most important features; there is still a great deal of structure they can not represent, however, and (for example) more rarely-explored parts of the space (including configurations that support human sacrifice, suicide, and castration) are not included.

As a second aid to visualization, we use Louvain clustering~\cite{blondel2008fast} to highlight groups of nodes that form well-defined communities (see Appendix Section~\ref{comm_detect} for details). Table~\ref{short_names} provides names for the religions labeled in the figure, and Table~\ref{distinctive_list} (and Appendix Table~\ref{distinctive_appendix}) provides a list of the most distinctive features of each group. The full list of group membership is provided in Appendix Table~\ref{observed_religions}.

\begin{table}
\centering
\begin{tabular}{lll}
\toprule
Group & Entry name (short) & Entry name (DRH) \\
\midrule
   Group 1 &        \href{https://religiondatabase.org/browse/654/#/}{Cistercians} &                            12th-13th c.\ Cistercians \\
    & \href{https://religiondatabase.org/browse/931/#/}{Jesuits in Britain} &          The Society of Jesus (Jesuits) in Britain \\
    & \href{https://religiondatabase.org/browse/738/#/}{Ancient Egyptian}   &                                   Ancient Egyptian \\
    & \href{https://religiondatabase.org/browse/1043/#/}{Islam in Aceh}  &                                      Islam in Aceh \\
   Group 2 &  \href{https://religiondatabase.org/browse/1311/#/}{Jehovah}  &                                Jehovah's Witnesses \\
    &  \href{https://religiondatabase.org/browse/879/#/}{Free Methodist}    &                              Free Methodist Church \\
    &   \href{https://religiondatabase.org/browse/984/#/}{Calvinism}        &                      Calvinism (Early/Reformation) \\
   Group 3 &  \href{https://religiondatabase.org/browse/230/#/}{Mesopotamia}        &                            Religion in Mesopotamia \\
    &   \href{https://religiondatabase.org/browse/1251/#/}{Tsonga}           &                                             Tsonga \\
    &     \href{https://religiondatabase.org/browse/534/#/}{Roman}              &                                Roman Imperial Cult \\
   Group 4 &  \href{https://religiondatabase.org/browse/1010/#/}{Pythagoreanism}     &                                     Pythagoreanism \\
    &      \href{https://religiondatabase.org/browse/1304/#/}{Peyote}        & Peyote Religion (Peyotism) \\
   Group 5 &    \href{https://religiondatabase.org/browse/769/#/}{Wogeo}              &                                              Wogeo \\
    &     \href{https://religiondatabase.org/browse/1511/#/}{Sokoto}          &                                             Sokoto \\
\bottomrule
\end{tabular}
\caption{Observed configurations labelled in Figure~\ref{config_networks}A; short names are hyperlinked to the original DRH entries.} \label{short_names} 
\end{table}

\begin{table}[ht]
    \centering
    \begin{tabular}{rll}
    \toprule
 Group & Color & Top distinctive practices \\
\midrule
    Group 1 & Grey & + monuments; + small-scale rituals; + scriptures \\
     Group 2 & Orange & -- small-scale rituals; -- grave goods; -- special corpse treatment \\
     Group 3 & Green & -- scriptures; + grave goods; -- monuments  \\
      Group 4 & Blue & -- supernatural monitoring, punishment; -- formal burials \\
     Group 5 & Pastel & -- supernatural monitoring, punishment; -- grave goods \\
     \bottomrule
    \end{tabular}
    \caption{Distinctive features of the five clusters in the landscape of Figure~\ref{config_networks}A; + indicates higher than average rates of ``yes''; --, higher than average rates of ``no''. For example, only 27\% of the probability mass in Group 3 is assigned to configurations that require small-scale rituals, compared to 89\% in the remainder of the sample. See Appendix Table~\ref{distinctive_appendix} for full list.} 
    \label{distinctive_list}
\end{table}
Group 1 (grey) is the largest by probability mass ($19\%$); it is characterized by monuments, small-scale rituals, and scriptures. Among others, this group contains Ancient Egyptian religions, many Islamic traditions, and officially-supported Catholic groups such as the Jesuits and Cistercians. 

Group 2 (orange) is the second-largest ($13\%$); it is characterized by generally lower probabilities of small-scale rituals, grave goods, and special corpse treatment, and includes demotic, charismatic, and reform traditions, including many Protestant groups such as the Southern Baptists, Jehovah's Witnesses, and Pentecostalism. Group 2, in contrast to Group 1, is more evenly weighted among its configurations; where Group 1 has a small number of peaks, Group 2 is more like a floodplain.

Group 3 (green) is the third largest by probability mass ($12\%$). It is characterized by a relative absence of written scripture, more grave goods, and fewer monuments. It includes many folk, traditional, indigenous, and ``pre-Axial''~\cite{bellah2017religion} pagan cultures, including the Roman Imperial Cults and Mesopotamian religions, as well as more recent revivals such as Exovedate and Postsocialist Mongolian Shamanism. It is also characterized,  compared to Groups 1 and 2, by a relative absence of moralizing ``big Gods''~\cite{whitehouse2022testing} who conduct supernatural monitoring and punishment.

Finally, Groups 4 (blue) and 5 (pastel) are much smaller ($0.75\%$ and $0.54\%$ probability, respectively). Group 4 includes Peyotism, Pythagoreanism, and the Temple of the Jedi Order; only two religions in our data (Wogeo and Sokoto) appear in Group 5. Both of these smaller communities are characterized by the absence of both supernatural monitoring and punishment.

While the landscape is inferred without reference to time, cultural evolution appears to have explored the landscape in a somewhat sequential fashion. These temporal effects include shifts from Group 3's Pre-Axial tribal and archaic religious cultures towards Group 1's later Axial religious cultures~\cite{bellah2017religion} and ``big Gods'' that co-evolved with large-scale complex societies~\cite{norenzayan2013big,SPICER2022100073}. Group 3 religions tend to be older than those in nearby Group 1, which has the highest concentration of religious cultures committed to a belief in high Gods, including many examples from the Abrahamic traditions. Group 2, in turn, includes popular developments out of Group 3 traditions into contemporary society, such as Pentecostalism, a sect established in the twentieth century, and rapidly becoming one of the largest Christian sub-groups~\cite{norenzayan2013big}.

The landscape reflects more than just a temporal sequence of social, economic, and material evolution, however. It also seems to capture the constraints of more permanent features of the human mind. While Group 1 includes many later ``solutions'' to the constraints found by Axial-age and ``Big Five'' religions, it also includes cases such as pre-Christian Ireland. Group 3, meanwhile, includes not just ancient pagan traditions, but modern revivals. Religions, in other words, may co-evolve with social context, but they also have to respect the psychological constraints on how we believe and keep faith, and may well wander back to earlier solutions~\cite{luhrmann2020god}.

\subsection{Focal Landscapes}
\label{focal}

Figure~\ref{config_networks}A provides an overview of how the constraints combine to make a landscape; a second way to explore is to map the nearby configurations of a particlular group. Among other things, this provides a grounded way to speculate on how a culture might evolve into the future, or where it might have come from---to ask, for example, which bits in a configuration might flip, and whether or not this would push the religion to a more probable configuration that is more able to satisfy the underlying constraints.

Figure~\ref{config_networks}B and~\ref{config_networks}C does this for two groups in our data, the (contemporary) Free Methodist Church and the (ancient) Roman Imperial Cult. In both cases, we take the group as the focal node, and include the 49 most probable nearby configurations, that differ in up to two answers from the focal case.

As seen in Figure~\ref{config_networks}B, the Free Methodist Church is situated at a local peak, and nearby configurations are of lower probability. Some of them appear in our data (\emph{e.g.}, the Southern Baptists, Pauline Christianity), but several are unoccupied. The highest probability configuration in the local region is Jehovah's Witnesses, two steps away. 

The Free Methodist Church does not require participation in large-scale rituals. A change in this attribute is their most probable reformation ($15\%$) and would place them in the same configuration as the Southern Baptists. Slightly less probable is the configuration in which the Free Methodists require participation in small-scale rituals ($10\%$). This change would result in the Free Methodists sharing a configuration with Pauline Christianity. Either change would take them closer to the Jehovah's Witnesses, but also would require them to go ``down'' through one of those two lower probability states.

In contrast to the Free Methodist Church, the Roman Imperial Cult (Figure~\ref{config_networks}C) sits in a valley, with several neighboring configurations of higher probability. The Cult satisfies the constraints better without its own distinct written language, than with (as was actually the case), and with scriptures rather than without. Loss of its own distinct language would shift it up to the Mesopotamia configuration, while acquiring scriptures would shift it up to the Achaemenid configuration.

\section{Discussion}

The main goal of this work has been to provide those in cultural evolution and sociophysics with new methods, and accompanying code, for inferring the landscapes beneath the incomplete data of the historical record. In addition to characterizing these methods through simulation, we have shown how they play out in a real-world example, drawn from the Database of Religious History. In the words of archaeologist David Hurst Thomas, ``it's not what you find, it's what you find out'', and we have endeavored to show how landscape models not only organize data from the field, but provide insight into the underlying laws and dynamics that can help explain it.

A key direction for future research is to consider how these methods might be extended to even larger configuration spaces. As the number of features considered rises higher and higher, so do the challenges; when $n$ goes from 20 to 100, for example, the number of parameters goes from 210 to more than 5000. To maintain the same level of accuracy would, generically, require the amount of data to rise by a similar factor---but this may not always be possible; in the final analysis, there are only a finite number of civilizations in human history.

A more creative solution to the problem is to go from the ``unrestricted'' Boltzmann machine case, where all $J_{ij}$s are (potentially) non-zero, to the ``restricted'' case, where some links are set to zero by the researcher ahead of time. In this case, the researcher sculpts a theory of constraints, restricting \emph{a priori} the ways in which features may interact and reducing the number of free parameters. Another solution is to connect nodes not to each other, but to a small number of hidden variables---``layers'', in the deep learning jargon. If there are $n$ features, and $m$ hidden nodes, then the total number of parameters, including local fields, is $n(m+1)$, which may make the problem tractable again. Hidden layers have proven to be particularly expressive; in the physics jargon, they are equivalent to how renormalization leads to higher-order interactions~\cite{mehta2014exact}. The original MPF paper~\cite{mpf} demonstrated the use of hidden nodes in this fashion, and the framework makes it possible to extend our Partial-MPF algorithm to these cases as well.

There are challenges in inference. There are equally compelling challenges in data gathering itself. The DRH is one example of the exciting resources coming on-line for researchers in the human sciences, but these sources bring complexities of interpretation in their wake. As discussed in Appendix Section~\ref{duplicate}, for example, drawing the boundaries between one group and another---in space, or time---is not a simple matter. This raises questions about how to properly combine the rich, qualitative data that comes from the field in ways that properly represent the diversity of human possibilities.



\funding{This work used the Extreme Science and Engineering Discovery Environment (XSEDE~\cite{towns2014xsede}), which is supported by National Science Foundation grant number ACI-1548562. Specifically, it used the Bridges-2 system~\cite{brown2021bridges}, which is supported by NSF award number ACI-1928147, at the Pittsburgh Supercomputing Center (PSC), under grant HUM220003. The DRH is funded by the John Templeton Foundation, Templeton Religious Trust, and Canada’s Social Sciences and Humanities Research Council (SSHRC). This work was supported in part by the Survival and Flourishing Fund.}

\dataavailability{Data and open-source code (incl. optimized C code {\tt mpf\_CMU}) for the methods and analysis described in this paper is available at \url{https://github.com/victor-m-p/humanities-glass}.} 

\acknowledgments{We thank the DRH team for discussions and data-sharing, and participants in the Santa Fe Institute workshop ``Coding the Past: The Challenges and Promise of Large-Scale Cultural Databases'' for discussions, that made this work possible.}

\conflictsofinterest{The authors declare no conflict of interest. The funders had no role in the design of the study; in the collection, analyses, or interpretation of data; in the writing of the manuscript; or in the decision to publish the~results.} 



\clearpage
\appendixtitles{yes} 
\appendixstart
\appendix

\section[\appendixname~\thesection]{Database of Religious History: Analysis and Data Considerations}

\subsection{Network Layout}\label{net_layout}
As briefly touched upon in section~\ref{visualize} it is mathematically impossible to faithfully represent the 20-dimensional hyper-cube landscape in a 2-dimensional layout. By only laying out a subset of the possible configurations (e.g. in Figure~\ref{config_networks}A, $150$ out of the total $2^{20}$ possible configurations) dimensionality reduction techniques can approximately compress the high-dimensional space into a low-dimensional spatial representation. We attempted approaches based on minimizing a global energy function (e.g. Multi-Dimensional Scaling, and similar approaches~\cite{gansner2004graph}), as well as force-directed placement algorithms (e.g. Fruchterman-Reingold heuristic~\cite{fruchterman1991graph}). We achieve the most appealing results following the latter approach, using the algorithm as implemented in Graphviz~\cite{ellson2001graphviz}. We stick with this approach for network layouts throughout, i.e., in all the plots shown in Figure~\ref{config_networks}, and in both plots shown in Figure~\ref{param_plot}. For the plots shown in Figure~\ref{config_networks} the layout uses only immediate neighbors (1 Hamming distance) and is unweighted. To create the spatial layout of nodes for Figure~\ref{param_plot} we threshold the connections (couplings) between nodes (Figure~\ref{param_plot}A), such that only connections with an absolute coupling value above $0.15$ are taken into account when running the force-directed algorithm. Figure~\ref{param_plot}B uses the layout obtained from Figure~\ref{param_plot}A to facilitate comparison.

\subsection{Community Detection}\label{comm_detect}

In Section~\ref{visualize} we report on five groups (communities) of configurations in the landscape of high-probability configurations. These are obtained using the Louvain community detection algorithm~\cite{blondel2008fast} as implemented in the Python package NetworkX~\cite{hagberg2008exploring}. The Louvain community detection algorithm is based on modularity optimization~\cite{blondel2008fast} and is a popular choice. The algorithm allows us to specify a ``resolution" parameter, which can bias the model towards fewer and larger communities ($\text{resolution} < 1$) or more and smaller communities ($\text{resolution} > 1$). The quality of the obtained communities can be assessed by the modularity which the algorithm is trying to optimize. While this is useful as an objective function of quality, the communities in our case are tools for interpretation, and do not represent a ground truth. We rely on human judgement, as well as objective measures, to decide on the proper (most meaningful) resolution. We obtain the most appealing results by biasing the algorithm slightly towards fewer and larger communities, and the five groups described in Section~\ref{visualize} were found using $\text{resolution} = 0.5$. In contrast, leaving the algorithm unbiased with $\text{resolution} = 1$ results in eight communities. As with the spatial layouts of the networks (see~\ref{net_layout}), we only provide connections (edges) between nodes (configurations) with Hamming distance 1 to the community detection algorithm. 

\subsection{Duplicate Entry Names}
\label{duplicate}

Entries in the Database of Religious History (DRH) all have a unique Entry ID, and the record has a name associated with it (Entry Name), but this name is not necessarily unique. In our sample of 407 religious cultures (Entry IDs) we have two religious cultures with the exact same Entry Name, Donatism and Roman Private Religion. In the case of Donatism we do in fact appear to have two overlapping entries from different experts, one which focus on Donatism from 311 CE to 427 CE, and one which focus on Donatism from 311 CE to 600 CE. In this one case, a legitimate argument can be made that it would be more appropriate to collapse this into one record about Donatism, using the flexibility of our MPF algorithm to weight potential disagreement. The Roman Private Religion case is different, with both entries submitted by the same expert, one describing the religious culture between 202 BCE and 44 BCE, and the other one describing the religious culture between 600 BCE and 202 BCE. In this case, the periods do not overlap, and it is reasonable to assume that the expert has made the decision to split the records based on differences between the two cultures that she has expert knowledge about. These cases raise a more general point about independence, and how we should treat partially overlapping cultures. 

The most extreme example from our curated subset of the DRH is the case of the Ancient Egyptian religions. We have a total of six different entries about Ancient Egyptian religions (e.g. early dynastic, first intermediate period, old kingdom, etc.). Since these cultures naturally overlap on most attributes, the model will consider Ancient Egyptian religions as very stable configurations (which is, in fact, not totally unreasonable). Four of these six different entries focusing on Ancient Egyptian religions have the same configuration, and this configuration is assigned the third highest probability mass in the landscape (annotated as ``Ancient Egyption" in figure~\ref{config_networks}A). Whether this is reasonable, or whether all of the Ancient Egyptian religions should be treated as one religious culture (and be weighted accordingly) is a difficult question. Because culture is fluid, and no culture is completely independent from other - past and present - cultures, it seems impossible to design a general decision rule for whether to consider two cultures meaningfully independent. In this paper, we have taken the records from the DRH at face value, and treated each unique Entry ID as its own religious culture. In the future, more sophisticated approaches should be pursued in collaboration with domain experts. 

\section[\appendixname~\thesection]{Database of Religious History: Religions and Question Codes used in this analysis}

\begin{table}[ht]
\centering
\caption{Distinctive community features}
\begin{tabular}{lllrrr}
\toprule
Group & Color & Question &  Avg. S &  Avg. O &  Diff \\
\midrule
   Group 1 &       Grey & Monumental religious architecture &               94.45 &               34.17 &              60.28 \\
   & &      Small-scale rituals required &               98.44 &               52.02 &              46.43 \\
   & &                        Scriptures &              100.00 &               55.05 &              44.95 \\
   Group 2 &     Orange &      Small-scale rituals required &               27.40 &               89.07 &             -61.67 \\
   & &                       Grave goods &               14.69 &               57.92 &             -43.23 \\
   & &     Special treatment for corpses &               48.65 &               87.15 &             -38.50 \\
   Group 3 &      Green &                        Scriptures &               23.96 &               91.91 &             -67.95 \\
   & &                       Grave goods &               85.58 &               30.71 &              54.87 \\
   & & Monumental religious architecture &               33.45 &               68.57 &             -35.12 \\
   Group 4 &       Blue &        Supernatural beings punish &                0.00 &               94.55 &             -94.55 \\
   & &                    Formal burials &                0.00 &               94.18 &             -94.18 \\
   & &   Supernatural monitoring present &                0.00 &               94.09 &             -94.09 \\
   Group 5 &     Pastel &        Supernatural beings punish &                0.00 &               94.11 &             -94.11 \\
   & &   Supernatural monitoring present &                0.00 &               93.65 &             -93.65 \\
   & &                       Grave goods &                0.00 &               45.98 &             -45.98 \\
\bottomrule
\end{tabular}
\caption{For each community we calculate the average possession of each of the 20 religious attributes. We weight this by the probability of each configuration in the community, and convert it to a percentage (Avg. S). We compare this to the average possession of the attribute across all configurations that are not in the community. Again we weight this by the probability of each configuration and convert it to a percentage (Avg. O). We calculate the difference (Diff), and show the three attributes for which each community differ most from the rest.}
\label{distinctive_appendix}
\end{table}

\begin{table}[ht]
\centering 
\begin{tabular}{r|l}
\toprule
 ID & Question (short; long) \\
\midrule
        1  &         Official political support  \\ 
 &                                                           Does the religion have official political support  \\ \hline 
         2  &                         Scriptures  \\ 
 &                                                                   Does the religious group have scriptures?  \\ \hline 
         3  &  Monumental religious architecture  \\ 
 &                                                               Is monumental religious architecture present?  \\ \hline 
         4  &            Spirit-body distinction  \\
\bottomrule
\end{tabular}
\caption{Question subset used in DRH analysis. ``Long'' question names, \emph{e.g.}, ``Does the religion have official political support", correspond to ``Related Question" as they appear in the DRH, besides ending characters which have been modified, e.g. ``Does the religious group have scriptures?" instead of ``Does the religious group have scriptures:" as it appears in the DRH. Short question names are used for convenience. ID column has been recoded to range from 1-20, and does not correspond to ``Related Question ID" in the DRH.}
\label{question_table}
\end{table}

\begin{longtable}{ p{.10\textwidth} | p{.10\textwidth} | p{.6\textwidth} | p{.10\textwidth}}
\toprule
  Group &  DRH ID &                                                                Entry Name &  Weight \\
\midrule
Group 1 &    1321 &                                                      Mourides (Muridiyya) &  100.00 \\
Group 1 &    1293 &                                The Zapotec or Ben 'Zaa (The Cloud People) &  100.00 \\
Group 1 &    1178 &                                               Modern Mystery School (MMS) &  100.00 \\
Group 1 &    1140 &                                               Early Medieval Confucianism &  100.00 \\
Group 1 &    1125 &                                      Ancient Egypt - the Ramesside Period &  100.00 \\
Group 1 &    1043 &                                                             Islam in Aceh &  100.00 \\
Group 1 &    1041 &                                                        Korean Catholicism &  100.00 \\
\end{longtable}

\begin{longtable}{ p{.10\textwidth} | p{0.9\textwidth}}
\toprule
 DRH ID &                                                                                                                            Entry Name \\
\midrule
     23 &                                                                                                                   Late Shang Religion \\
    173 &                                                                                                                Johannine Christianity \\
    174 &                                                                                                        Matthew-James-Didache Movement \\
    190 &                                                                                                    Sri Lankan Buddhism (1948-Present) \\
    204 &                                                                                                                      Han Confucianism \\
    217 &                                                                                                                Roman private religion \\
    222 &                                                                                                             Late Classic Lowland Maya \\
\end{longtable}


\begin{adjustwidth}{-\extralength}{0cm}

\clearpage
\reftitle{References}


\bibliography{main}




\end{adjustwidth}
\end{document}